\documentclass[journal]{IEEEtran}
\usepackage{times}
\usepackage{epsfig}
\usepackage{graphicx}
\usepackage{amsmath}
\usepackage{amssymb}
\usepackage{mathtools}
\usepackage{cite}
\usepackage{multirow}
\usepackage{pifont}
\usepackage{booktabs}
\usepackage{stfloats}
\usepackage[para]{threeparttable}
\usepackage{nicefrac}
\usepackage[switch]{lineno}
\newcommand{\X}{\mathbf{x}}
\newcommand{\Y}{\mathbf{y}}
\newcommand{\Z}{\mathbf{z}}

\newcommand{\etal}{\textit{et al}. }
\newcommand*\xbar[1]{%
   \hbox{%
     \vbox{%
       \hrule height 0.2pt % The actual bar
       \kern0.5ex%         % Distance between bar and symbol
       \hbox{%
         \kern-0.1em%      % Shortening on the left side
         \ensuremath{#1}%
         \kern-0.1em%      % Shortening on the right side
       }%
     }%
   }%
} 
\newcommand{\RNum}[1]{\uppercase\expandafter{\romannumeral #1\relax}}
\begin{document}
%\linenumbers
% paper title
% Titles are generally capitalized except for words such as a, an, and, as,
% at, but, by, for, in, nor, of, on, or, the, to and up, which are usually
% not capitalized unless they are the first or last word of the title.
% Linebreaks \\ can be used within to get better formatting as desired.
% Do not put math or special symbols in the title.
\title{A Resolution Enhancement Plug-in for Deformable Registration of Medical Images}
%
%
% author names and IEEE memberships
% note positions of commas and nonbreaking spaces ( ~ ) LaTeX will not break
% a structure at a ~ so this keeps an author's name from being broken across
% two lines.
% use \thanks{} to gain access to the first footnote area
% a separate \thanks must be used for each paragraph as LaTeX2e's \thanks
% was not built to handle multiple paragraphs
%

\author{Kaicong Sun, Sven Simon% <-this % stops a space
\thanks{K. Sun and S. Simon are with the Institute for Parallel and Distributed Systems, University of Stuttgart, Universit\"{a}tsstra\ss e 38, 70569 Stuttgart, Germany. (e-mail: kaicong.sun@ipvs.uni-stuttgart.de). This work was supported by the Federal Ministry of Education and Research (BMBF, Germany) under the grand No. 105M18VSA.}% <-this % stops a space
\thanks{Manuscript received on xxx December, 2021; revised xxx, 2022.}}

\maketitle

\begin{abstract}
Image registration is a fundamental task for medical imaging. Resampling of the intensity values is required during registration and better spatial resolution with finer and sharper structures can improve the resampling performance and hence the registration accuracy. %Spatial resolution is one of the determinant quality indicators of imaging systems which characterizes the finest structure the system can distinguish. 
Super-resolution (SR) is an algorithmic technique targeting at spatial resolution enhancement which can achieve an image resolution beyond the hardware limitation. In this work, we consider SR as a preprocessing technique and present a CNN-based resolution enhancement module (REM) which can be easily plugged into the registration network in a cascaded manner. Different residual schemes and network configurations of REM are investigated to obtain an effective architecture design of REM. In fact, REM is not confined to image registration, it can also be straightforwardly integrated into other vision tasks for enhanced resolution. The proposed REM is thoroughly evaluated for deformable registration on medical images quantitatively and qualitatively at different upscaling factors. Experiments on LPBA40 brain MRI dataset demonstrate that REM not only improves the registration accuracy, especially when the input images suffer from degraded spatial resolution, but also generates resolution enhanced images which can be exploited for successive diagnosis.   
\end{abstract}


\IEEEpeerreviewmaketitle

\section{Introduction}
\IEEEPARstart{D}{eformable} image registration is a technique aiming for establishing dense spatial correspondence
within an image pair based on structure or intensity similarity~\cite{survey1,survey2,survey4,survey5}. It is a fundamental task for many medical applications such as detecting temporal anatomical changes, analyzing variability across populations, and multi-modality fusion. Most of the conventional deformable registration algorithms deal with an objective function $J$ formulated as
\begin{linenomath}
\begin{equation}\label{Objective}
J = D(F(\X), \phi\circ M(\X)) + \lambda R(\phi).
\end{equation}
\end{linenomath}
The data fidelity term $D$ measures the alignment between the fixed image $F$ and the transformed moving image $\phi\circ M$ based on such as L2 error norm, mutual information, and cross correlation. The dense deformation field $\phi$ maps the coordinates of the fixed image $F$ to the coordinates of the moving image $M$ and the symbol $\circ$ acts as resampling of the moving image $M$ with $\X\in \mathbb{R}^3$ in the 3D Cartesian coordinate system. $R$ indicates the regularization term which imposes some specified constraint on the deformation field such as smoothness. 

Different from most of the registration methods which concern either the biomedical constraints such as diffeomorphism~\cite{Beg2, Avants, Ashburner2} or computational acceleration~\cite{dalca, FDRN,VoxelMorph}, we look into the registration problem from another perspective. Since resampling of the intensity values of the moving image is required during image registration, images containing finer and sharper structures in larger image grid would push the registration accuracy towards a higher limit. To this end, super-resolution (SR), which is an algorithm-based technique dedicated to spatial resolution enhancement, is leveraged as a preprocessing step to pave the way for a better registration performance especially when the input image pair suffers from degraded spatial resolution. 

Due to the success of deep learning, in this work we propose a CNN-based resolution enhancement module (REM) using residual learning. It can be easily plugged into the registration network in a cascaded manner. Note that the proposed cascaded framework is not confined to image registration, REM can also be applied to other vision tasks such as image segmentation, object detection, and scene recognition either by employing some coupling techniques such as auxiliary loss or by a straightforward cascade. The contributions of our work are summarized in the following:
\begin{itemize}
\item A lightweight resolution enhancement module (REM) based on residual CNN is presented which can be easily integrated into vision networks as preprocessing for augmented resolution. To achieve an efficient architecture design, different residual schemes and network configurations of REM are investigated.
\item An auxiliary loss is introduced into the cascaded registration network to empower multi-hierarchical supervision and strengthen the fidelity of the output.
\item The proposed REM and the cascaded framework are thoroughly evaluated based on two registration networks FDRN~\cite{FDRN} and VoxelMorph~\cite{VoxelMorph} for different upscaling factors quantitatively and qualitatively. Experiments show that REM improves not only the registration accuracy but also produces super-resolved images which can be utilized for further analysis.
\end{itemize}

\section{Related Work}

\subsection{Deformable Image Registration}

In the literature, deformable registration can be classified into two categories based on the model of the deformation field, either by the use of displacement vector field (DVF) d(x) or the velocity vector field $v(\X, t)$. The former category models the spatial transformation as a linear combination of the identity transform $\X$ and the DVF: $\phi = \X+d(\X)$ with $d(\X)$ representing the spatial offsets between the corresponding voxels in the fixed and the moving images~\cite{Bajcsy, Gee, Davatzikos, Rueckert, Kybic, Sdika, Thirion, FDRN, Balakrishnan, VoxelMorph}. Generally, the DVF-based deformation model might have difficulty in dealing with registration of large deformations as the transformation is linearized around the coordinate system of the moving image. While the latter one concerns the invertibility of the transformation~\cite{Beg1, Beg2, Avants, Ashburner2, dalca} by involving biomedical constraints such as diffeomorphism, topology preservation, inverse consistency, and symmetry on the deformation field. Specially, deformable registration is considered as a variational problem and $\phi$ is formulated as a definite integral of the velocity vector field $v(\X, t)$. However, despite of the categories, most of these work address the registration problem either from the perspective of applying different biomedical constraints or from the aspect of computational acceleration, we look into the registration problem from the view of resolution enhancement, which is effective and promising especially when the input images suffer from low spatial resolution.  

\subsection{Image Super-Resolution}
In the last few decades, image SR has been intensively investigated from frequency domain~\cite{Huang, Kim2} to image domain~\cite{BTV,Elad,Hardie,IRWSR,MPG,MPGBSWTV,FLMISR,SRCNN,SRGAN,VDSR,EDSR,DPSR,ESRGAN}, from conventional optimization-based methods~\cite{BTV,Elad,Hardie,IRWSR,MPG,MPGBSWTV,FLMISR} to more recently emerged deep learning-based approaches~\cite{SRCNN,SRGAN,VDSR,EDSR,DPSR,ESRGAN}. Usually, the conventional SR methods perform iterative optimization and require multiple low-resolution (LR) images of the same scene captured under certain geometric offsets to restore the high-resolution (HR) counterpart. On the contrary, the learning-based approaches ``hallucinate''  the high-frequency details from a single LR input exclusively based on the large-scale training datasets. Comparing to the learning-based ones, the performance of the traditional model-based algorithms relies more on the sophisticated objective function and the quality of the available inputs. Besides, the traditional optimization-based methods require usually more computation time due to the iterative update scheme. While the learning-based approaches benefit greatly from the assembled massive external examples in the training datasets and also the GPU accelerated platform. Therefore, the deep learning-based SR reconstruction is more favored in recent years in the SR community.

\subsection{Resolution Enhancement for Other Tasks}
SR is favored in most of the vision tasks by increasing the image resolution and improving the image quality. In~\cite{DaiD}, Dai~\etal quantitatively demonstrate the effectiveness of SR as preprocessing on other vision tasks such as edge detection, semantic image segmentation, digit recognition, and scene recognition. Some other work take advantage of the super-resolved features for specific vision tasks such as semantic segmentation~\cite{dualsr}, face recognition~\cite{dualface} and pedestrian identification~\cite{srreid}. However, these methods require a highly specialized SR approach for the individual task which makes the embedding of SR less convenient. In this work, a general purpose handy network REM is proposed which can be easily integrated into other tasks in a cascaded manner. The cascaded framework is evaluated for deformable image registration and we demonstrate that REM not only improves the registration accuracy but also generates resolution enhanced images in the meanwhile which can be utilized for successive diagnosis. 

\begin{figure}
\centering
	\includegraphics[width=0.46\textwidth]{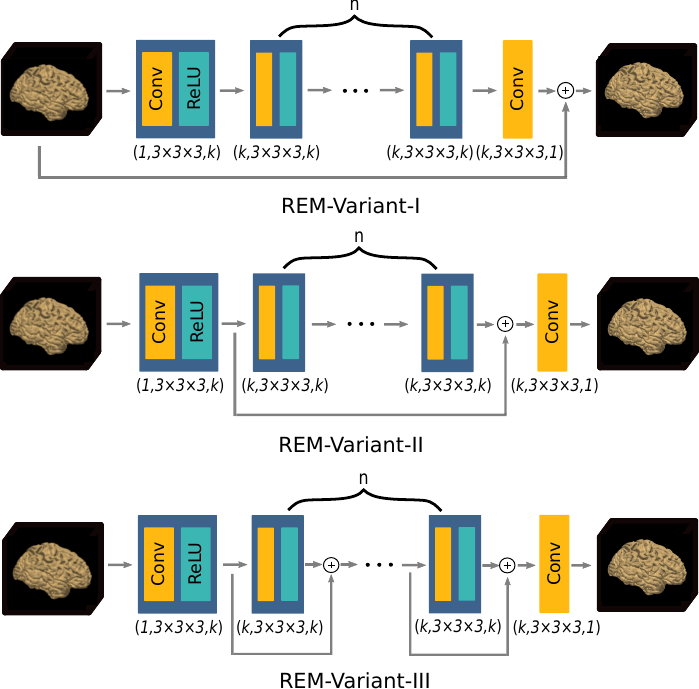}
	\caption{Different variants of the resolution enhancement module (REM).}
	\label{fig:REM}
\end{figure}

\begin{table*}[hb!]
	\centering
	\setlength{\tabcolsep}{8pt}
	\caption{Evaluation of different REM variants in PSNR, SSIM, number of parameters, and required GPU memory. The notations $k16n8$ and $k16n16$ indicate the configuration of $k=16, n=8$ and $k=16, n=16$, respectively.}
	\label{tab:REMVariant}
	\begin{threeparttable}
	\begin{tabular}{c c c c c c c}
		\toprule
		 &\multicolumn{2}{c}{REM-Variant-I} &\multicolumn{2}{c}{REM-Variant-II} &\multicolumn{2}{c}{REM-Variant-III} \\ \midrule
		  &\multicolumn{1}{|c}{$k16n8$} &$k16n16$ &\multicolumn{1}{|c}{$k16n8$} &$k16n16$ &\multicolumn{1}{|c}{$k16n8$} &$k16n16$ \\ \midrule
	PSNR/SSIM &\multicolumn{1}{|c}{45.43/0.9956} & 45.57/0.9958 &\multicolumn{1}{|c}{45.38/0.9952} & 45.51/0.9957 &\multicolumn{1}{|c}{45.38/0.9955} &45.34/0.9955  \\ \midrule
		\#Parameters [K] &\multicolumn{1}{|c}{56.3} &111.7 &\multicolumn{1}{|c}{56.3} &111.7 &\multicolumn{1}{|c}{56.3} &111.7 \\ \midrule
		Train/Test Memory [GB] &\multicolumn{1}{|c}{1.4/2.4} &2.0/2.4 &\multicolumn{1}{|c}{1.4/2.4} &2.0/2.4 &\multicolumn{1}{|c}{1.4/2.4} &2.0/2.4\\\bottomrule
	\end{tabular}
	\end{threeparttable}
\end{table*}

\begin{figure*}
\centering
	\includegraphics[width=0.95\textwidth]{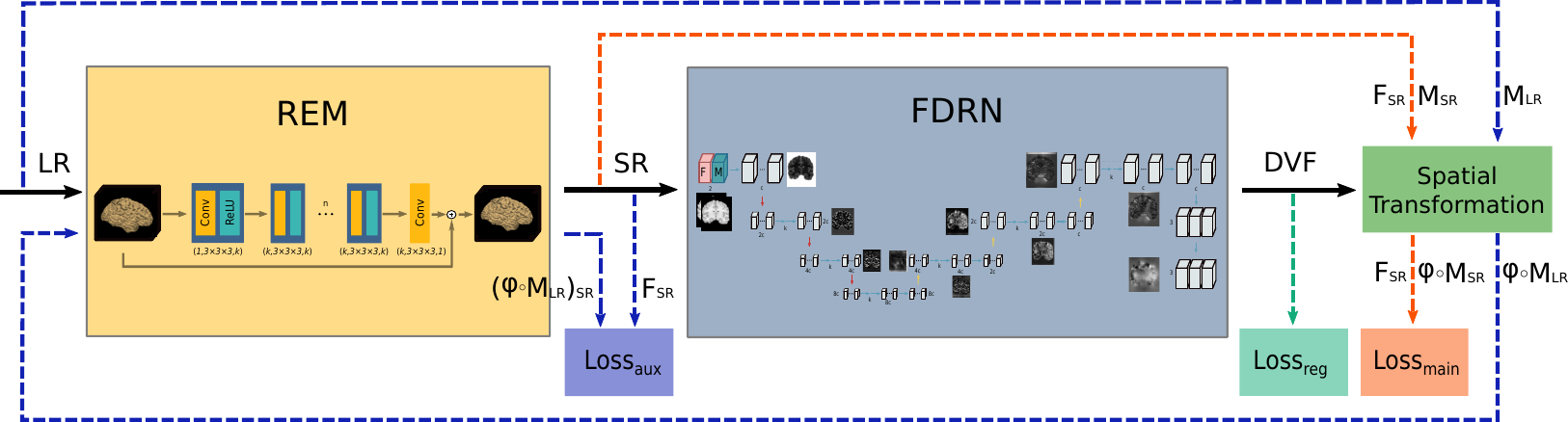}
	\caption{Cascaded registration network ReFDRN. The colored dotted lines represent the data flow of the different loss components. (Best viewed in color)}
	\label{fig:ReFDRN}
\end{figure*}

\section{Resolution Enhancement Module and Cascaded Network}
\subsection{Resolution Enhancement Module}
Deep learning brings up a new generation of SR. Since the pioneer work SRCNN~\cite{SRCNN}, learning-based SR has been intensively investigated from different aspects such as perceptual quality and extreme high upscaling factor. %of different upscaling factors, perceptual quality~\cite{zhang2020ntire, lugmayr2020ntire} in the last few years . 
Many representative methods~\cite{VDSR, SRGAN, SRDenseNet, EDSR, ESRGAN, rcan, DPSR} advance the state-of-the-art by adopting residual learning~\cite{VDSR}, dense connection~\cite{SRDenseNet}, GAN~\cite{SRGAN,ESRGAN}, enhanced depth~\cite{EDSR}, attention~\cite{rcan}, plug-and-play~\cite{DPSR} to count a few. In fact, all the aforementioned models make use of the extremely distinguished capability of CNN for feature extraction. We intend to combine a general purpose CNN-based resolution enhancement network with the registration network and propose a cascaded framework for enhancing the performance of image registration. Therefore, the resolution enhancement network acts as an auxiliary module and is desired to be a lightweight plug-in which should be simple yet effective. 

We firstly demonstrate three potential candidates of REM with different residual architectures schematically illustrated in Fig.~\ref{fig:REM}. REM-Variant-I adopts the additive skip connection from end-to-end on the image domain such that the network purely learns the residual high frequency components. REM-Variant-II applies the residual scheme on the feature domain and hence the intermediate convolutional layers learn the residual features. REM-Variant-III intends to aggressively exploit skip connection on each convolutional layer. In order to circumvent the architecture with a handcrafted upscaling factor, REM processes the trilinear upsampled image as the input. The neat structure contains successive 3D convolutional layers with Rectified Linear Unit (ReLU) inbetween and the convolution kernel size is set as $3\times 3\times 3$. Residual learning is adopted to improve the learning efficiency. The amount of filters in the convolutional layers is indicated by $k$ and the number of intermediate convolutional layers is denoted by $n$. To hold a compact design, based on empirical observations we set $k\leq 16, n\leq 16$ since larger $k$ and $n$ would either bring limited performance gain for brain MR images or cause overfitting. Due to the fact that REM does not alter the dimension of the input, it can be straightforwardly embedded into different vision tasks.

In Table~\ref{tab:REMVariant}, we summarize the performance of different variants of REM for the upscaling of 2$\times$ in PSNR, SSIM, number of parameters, and required memory. It turns out that Variant-I has the best performance in PSNR and SSIM comparing to the other two candidates even though they have the same number of network parameters. Variant-II achieves the second best but it requires more training time than Variant-I since Variant-I learns only the residual image. Although Variant-III exploits the residual scheme and is supposed to support more effective learning for networks with larger depth, it seems $k16n16$ does not perform better than $k16n8$. The reason might be that Variant-III is easier to be trapped at local optima. It is necessary to mention that for training, we take the patch size of 64$\times$64$\times$64,  while during testing the whole image of size 160$\times$208$\times$176 is fed into the network. That is the reason why training requires less memory than testing in the table. If we set the patch size as the whole image for training, $16k8n$ needs 8.1GB GPU RAM and $16k16n$ runs out of the 11GB memory in our GPU. In this work, we select REM-Variant-I as the architecture of REM. 

\subsection{Cascaded Network}

Combining REM with the registration network, the architecture of the proposed cascaded network is illustrated in Fig.~\ref{fig:ReFDRN}. For demonstration purpose, we choose FDRN~\cite{FDRN} as the deformable registration network and the cascaded network is denoted as ReFDRN. In fact, FDRN can be easily replaced by other ones without additional modifications. Particularly, ReFDRN takes LR images as the input and outputs the corresponding SR images and the DVF. The dotted lines in different colors denote the dataflow of the individual component of the loss function. %as formulated in \eqref{eq:ReFDRNLoss}. 
Mathematically, the reconstructed SR image $\Y$ and the DVF $\Z$ can be expressed as
\begin{equation}
\begin{split}
\Y &= REM(\X),~ \X\in\mathbb{R}^{2\times 1\times L\times W\times H}, \Y\in\mathbb{R}^{2\times 1\times L\times W\times H} \\
\Z &= FDRN(\tilde{\Y}),~ \tilde{\Y}\in\mathbb{R}^{1\times 2\times L\times W\times H}, \Z\in\mathbb{R}^{1\times 3\times L\times W\times H}  
\end{split}
\end{equation}
where $\tilde{\Y}$ is the rearrangement of $\Y$ by switching the batch and the channel dimension since REM needs to treat both input images individually, while FDRN considers them as an image pair and process them in one shot. The overall loss consists of three components as formulated in \eqref{eq:ReFDRNLoss}: the main loss $L_{main}$, the regularization term $L_{reg}$, and the introduced auxiliary loss $L_{aux}$. The weighting parameters $\lambda_1$ and $\lambda_2$ are scalars and are adopted to tradeoff between the main loss and the rest ones during the training.

\begin{equation}
\label{eq:ReFDRNLoss}
L_{total} = L_{main} + \lambda_1 L_{aux} + \lambda_2 L_{reg}
\end{equation}

In particular, the main loss $L_{main}$ focuses on the original task, i.e., image registration and in FDRN~\cite{FDRN}, the local normalized cross correlation (LNCC) is adopted. The difference between ReFDRN and FDRN in $L_{main}$ is that instead of processing on raw input images, ReFDRN performs registration on the super-resolved SR images as formulated in \eqref{eq:Lossmain}: 

\begin{equation}
\label{eq:Lossmain}
L_{main} = -LNCC\big(\phi\circ REM(M_{LR}),REM(F_{LR})\big).
\end{equation}

The regularization term is taken over from FDRN to impose smoothness constraint on the estimated DVF expressed as  
\begin{linenomath}
\begin{equation}\label{eq:Lossreg}
L_{reg}=\sum\limits_{S_m}||z-S_m z||^2_2,
\end{equation}
\end{linenomath}
where $S_m$ indicates the shifting operator along $(u,v,w)$ direction described by vector $m$ with $m = \{(u,v,w)~|~u,v,w\in\{0,1\}\}$ and $||\cdot||^2_2$ represents the L2-norm.

In order to strengthen the fidelity of the deformation field $\phi$, an auxiliary loss is introduced as below:  
\begin{equation}
\label{eq:Lossaux}
L_{aux} = f_{Huber}\big(REM(\phi\circ M_{LR}),REM(F_{LR})\big).
\end{equation}
The resampling is directly performed on the LR image to achieve a multi-hierarchical supervision. To complement the correlation-based main loss, an intensity-based Huber loss is utilized. We will demonstrate the effectiveness of the proposed auxiliary loss in Section~\ref{sec:evalauxiliary}. 

It is worthy noting that the auxiliary loss presented above is devised for unsupervised image registration to facilitate the coupling of the cascade and in fact, for other tasks if a similar design of the auxiliary loss is not available, the two networks can be straightforwardly connected by feeding the output of REM into the following one.

\section{Experiments and Results}
In this section, we firstly evaluate different configurations of REM (Variant-I) and demonstrate the performance of REM quantitatively in PSNR and SSIM and qualitatively in visual perception for upscalings of 2$\times$ and 4$\times$. We then evaluate the registration performance of the cascaded network. In addition to FDRN, we employ another representative learning-based registration network VoxelMorph~\cite{VoxelMorph} to validate the effectiveness of REM. As assessment metrics of image registration, Dice score and NCC are adopted. Finally, the effectiveness of the auxiliary loss on the registration performance of the cascaded network is studied and illustrated.
 
\begin{figure}
\centering
	\includegraphics[width=0.45\textwidth]{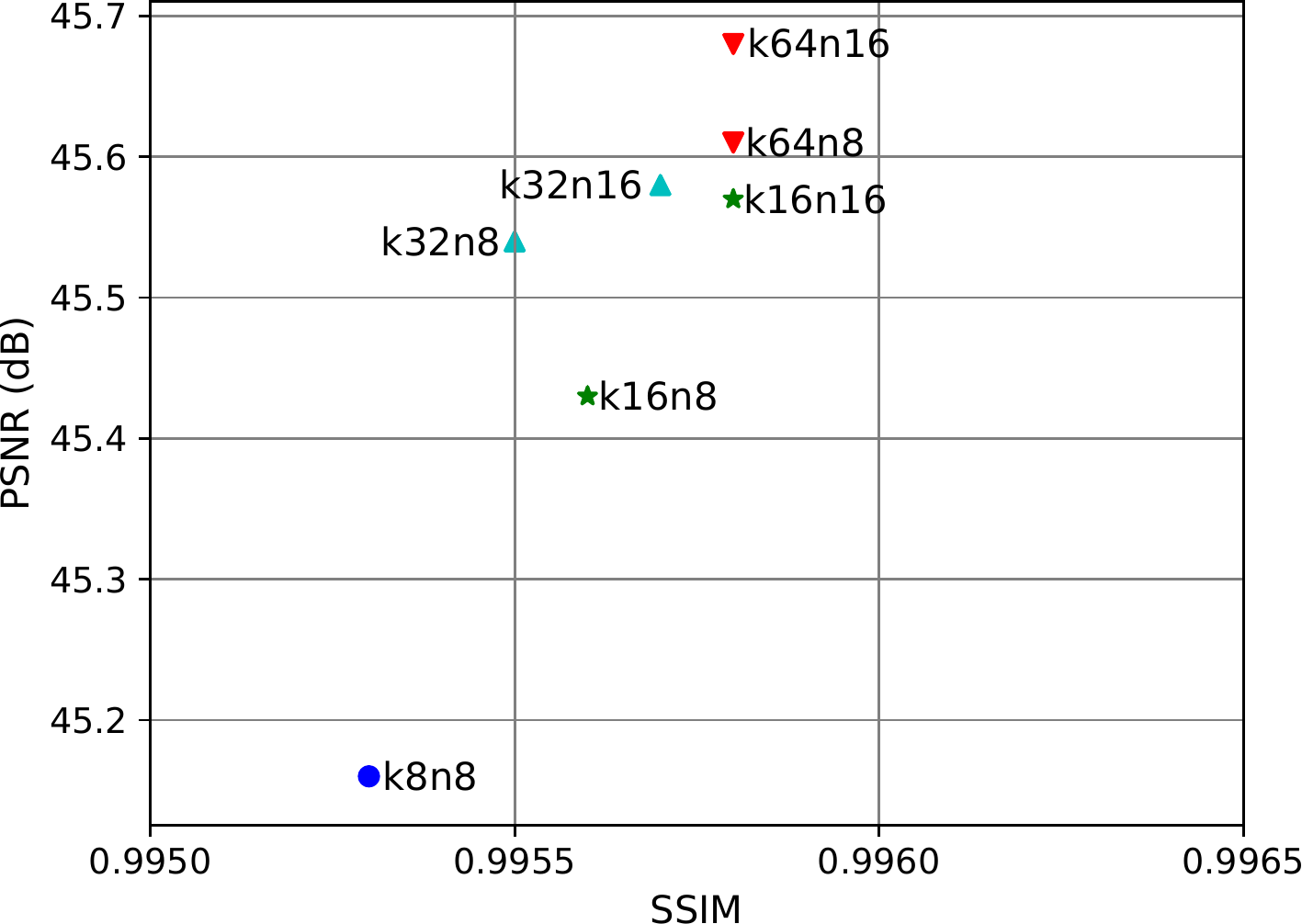}
	%\caption{$IoU: NN = 0.986411, Bic = 0.996811, SR = 0.999634$.}
	\caption{Performance of different configurations of REM in PSNR and SSIM on LPBA40. Different markers represent the configurations with different channel dimensions.}
	\label{fig:REMPerformance}
\end{figure}

\subsection{Dataset}\label{subsec:dataset2}
In this work, we have utilized the LPBA40 dataset~\cite{LPBA40} which contains brain MR images of 40 neurologically intact nonepileptic subjects with segmentation labels for 56 brain regions. All of these MR images were firstly registered to the Montreal Neurological Institute (MNI) space using affine transformation based on the ICBM152 template~\cite{unbiased}. The registered images were then cropped to the size of $160\times208\times176$. We have split the 40 cropped images into 30, 4, and 6 for training, validation, and testing, respectively. REM and the registration network were all trained based on LPBA40. For REM, the cropped MR images were considered as the ground truth (GT) and they were trilinear downscaled by factors of $\nicefrac{1}{2}$ and $\nicefrac{1}{4}$ which were taken as the LR inputs. The experimental results of REM and ReFDRN are demonstrated respectively in Section~\ref{sec:evalREM} and Section~\ref{sec:evalcascade}.

\subsection{Training Details}
\subsubsection{Training of REM}
REM was implemented with a Pytorch backend and trained on LPBA40 dataset based on Huber loss in a supervised manner. The mini-batch size was set as two and patch size was chosen as 64$\times$64$\times$64. Adam was used as the optimizer and the learning rate was initialized as 0.001 and multiplied by 0.95 every 200 iterations until decreased to 0.0001 over 1000 epochs. The experiments were performed on the NVIDIA GeForce GTX 1080 Ti GPU with 11GB GDDR5X and the Intel(R) Xeon(R) E5-2650 v2 CPU.
\subsubsection{Training of Cascaded Registration Network}
The pretrained REM was integrated into FDRN and the parameters of REM were frozen to preserve the fidelity of the super-resolved images and meanwhile ease the training of FDRN. To avoid missing of the corresponding voxels in the image pair during registration, a patch size of full image $160\times208\times176$ was employed and due to the memory limitation, a mini-batch size of one was used. Following the work of FDRN~\cite{FDRN}, the initial learning rate was set as 0.002 and multiplied by 0.9 every 1000 iterations until decreased to 0.0001 over 70 epochs. The weighting parameters were chosen as $\lambda_1=10$ for the auxiliary loss and $\lambda_2=1\times10^{-8}$ for the regularization. It is worth to mention that the abovementioned hyperparameters were tuned in a trial-and-error manner on the validation dataset and the ones generating the best Dice score were selected.

\subsection{Evaluation of Resolution Enhancement Module}\label{sec:evalREM}

\begin{figure*}
\centering
	\includegraphics[width=0.92\textwidth]{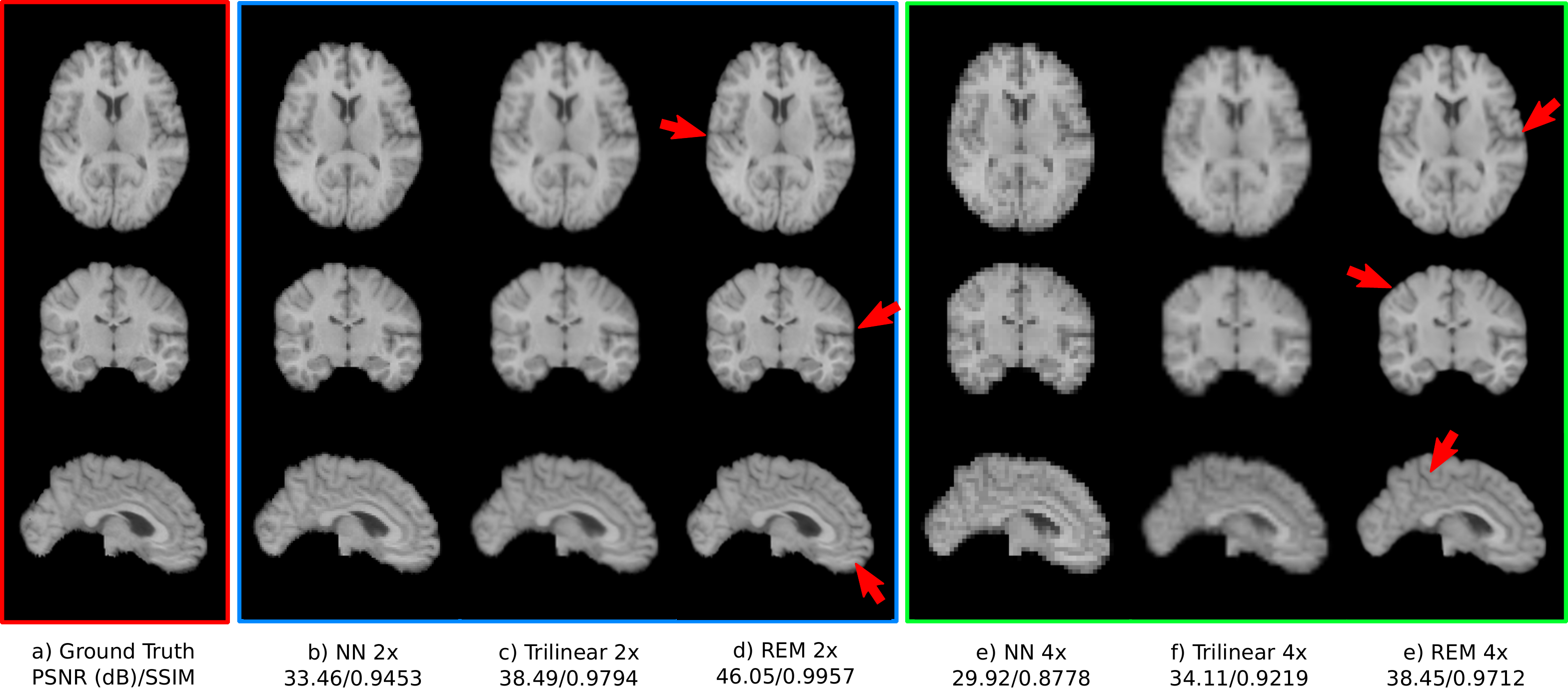}
	\caption{Evaluation of REM with the configuration of $k16n8$ for upscaling factors of 2$\times$ and 4$\times$ visually and quantitatively on the LPBA40 dataset.}
	\label{fig:REMVisual}
\end{figure*}

To figure out the network hyperparameters, a wide parameter sweep of the configurations of REM was performed. We set the channel number as $k=8, 16, 32, 64$ and the amount of intermediate convolutional layers as $n=8, 16$. The performance of all the investigated configurations is illustrated in Fig.~\ref{fig:REMPerformance}. It is shown that when $k>16$, the benefit margin of PSNR and SSIM is not evident and sometimes even reduced due to overfitting. $k=16$ performs certainly better than $k=8$. In fact, we have also conducted $n=32$ and it was severely overfitted. Besides evaluation of registration accuracy, we summarize the number of parameters and the required inference memory of all the configurations in Table~\ref{tab:REM_performance}. We can observe that the memory footprint during inference is not related to the number of layers but only to the amount of channels. Comparing to other SR networks such as EDSR~\cite{EDSR} which has millions of parameters although only for 2D SR, our lightweight 3D REM $k16n8$ only has about 50K parameters. In the latter experiments, we chose $k=16, n=8$ as the configuration of REM to balance the SR performance and the memory usage. Depending on the specifications and applications, one can choose the more appropriate configuration.

\begin{table}
\centering
\begin{threeparttable}[h]
\setlength{\tabcolsep}{2pt}
\caption{Number of network parameters and required inference memory of different REM configurations.}
 \label{tab:REM_performance}
\begin{tabular}{c c c c c c c c}
    \toprule
    Models & $k8n8$ & $k16n8$& $k16n16$ & $k32n8$&$k32n16$&$k64n8$&$k64n16$\\ \midrule
	\#Parameters [K] & 14.3&56.3&111.7& 223.2& 444.6& 888.8 &1774.0\\  	
	Memory [GB]& 1.7&2.4&2.4& 3.8& 3.8& 6.7 &6.7\\ 	
	\bottomrule 
   \end{tabular}
    \end{threeparttable}
\end{table}

\begin{table*}[hb!]
	\centering
	\setlength{\tabcolsep}{6pt}
	\caption{Performance evaluation of cascaded registration in Dice and NCC. The subscript $\downarrow\uparrow$ denotes trilinear downsampling followed by trilinear upsampling of the same scale and $\downarrow$ indicates downsampling without upsampling.}
	\label{tab:ReFDRN-ReVoxelMorph}
	\begin{threeparttable}
	\begin{tabular}{c c c c c c c c c}
		\toprule
		 Scales & Metrics &Affine$_{\downarrow}$ &FDRN$_{\downarrow}$ &VoxelMorph$_{\downarrow}$ &FDRN$_{\downarrow\uparrow}$ &ReFDRN$_{\downarrow\uparrow}$&VoxelMorph$_{\downarrow\uparrow}$ &ReVoxelMorph$_{\downarrow\uparrow}$ \\ \midrule
		1$\times$ &Dice/NCC &0.6079/0.9506 &0.6810/0.9976 &0.6746/0.9973&0.6810/0.9976 &--/-- &0.6746/0.9973 &--/--  \\ \midrule
		2$\times$ &Dice/NCC&0.6001/0.9511 &0.6369/0.9923 &0.6350/0.9920 &0.6797/0.9963  &0.6796/0.9977 &0.6722/0.9960 &0.6741/0.9970  \\ \midrule
		4$\times$ &Dice/NCC&0.5546/0.9396 &0.5536/0.9693 &0.5551/0.9695 &0.6676/0.9920 &0.6736/0.9962 &0.6593/0.9916 &0.6676/0.9932 \\ \bottomrule
	\end{tabular}
	\end{threeparttable}
\end{table*}

In Fig.~\ref{fig:REMVisual}, visual evaluation of REM ($k16n8$) for the upscalings of 2$\times$ (marked by blue rectangle) and 4$\times$ (marked by green rectangle) on LPBA40 dataset is demonstrated. Rows from top to bottom represent axial, coronal, and sagittal view, respectively. The very left column is the high-resolution GT and the results of nearest neighbor (NN) and trilinear interpolation are exhibited for comparison. It is shown that REM improves the visual quality significantly by generating sharper contours and providing better visibility of the detailed structures as marked by the arrows. Besides the visual assessment, quantitative evaluation in PSNR and SSIM is also depicted. Comparing to the naive interpolation-based upsamplings, REM achieves significantly better performance for both upscaling factors which coincides with the visual perception.

\iffalse
\begin{table*}[hb!]
	\centering
	\setlength{\tabcolsep}{5pt}
	\caption{Summary of performance evaluation in Dice and NCC. The subscript $\downarrow\uparrow$ denotes trilinear downsampling the input followed by the trilinear upsampling of the same scale and $\downarrow$ indicates NN downsampling without upsampling.}
	\label{tab:ReFDRN-ReVoxelMorph}
	\begin{threeparttable}
	\begin{tabular}{c c c c c c c c c}
		\toprule
		 Scales & Metrics &Affine$_{\downarrow}$ &FDRN$_{\downarrow}$ &VoxelMorph$_{\downarrow}$ &FDRN$_{\downarrow\uparrow}$ &ReFDRN$_{\downarrow\uparrow}$&VoxelMorph$_{\downarrow\uparrow}$ &ReVoxelMorph$_{\downarrow\uparrow}$ \\ \midrule
		\multirow{2}{*}{1$\times$} &Dice&0.6079 &0.6810 & 0.6746&0.6810 &-- &0.6746 &--  \\
								   &NCC&0.9506 &0.9976 & 0.9973& 0.9976 &-- &0.9973 &-- \\ \midrule
		\multirow{2}{*}{2$\times$} &Dice&0.6001 &0.6369 &0.6350 &0.6797 &0.6796 &0.6722 &0.6741  \\ 
								   &NCC&0.9511 &0.9923 &0.9920 &0.9963 & 0.9977 &0.9960 &0.9970 \\ \midrule
		\multirow{2}{*}{4$\times$} &Dice&0.5546 &0.5536 &0.5551 &0.6676 &0.6736 &0.6593 &0.6676 \\
								   &NCC&0.9396 &0.9693 &0.9695 &0.9920 &0.9962 &0.9916 &0.9932\\ \bottomrule
	\end{tabular}
	\end{threeparttable}
\end{table*}
\fi

\subsection{Evaluation of Cascaded Registration Network}\label{sec:evalcascade}

Since the proposed REM has shown great performance gain for spatial resolution enhancement as depicted in Section~\ref{sec:evalREM}, how does the improved image quality influence the performance of image registration? We plugged REM into two registration networks FDRN~\cite{FDRN} and VoxelMorph~\cite{VoxelMorph} following the cascaded framework as depicted in Fig.~\ref{fig:ReFDRN} and the cascaded networks are denoted as ReFDRN and ReVoxelMorph, respectively. To prepare the LR images, the GT images were firstly downscaled by trilinear interpolation with a factor of $\nicefrac{1}{2}$. The image dimension was then restored by a trilinear upscaling of factor 2$\times$. The trilinear upscaled images were passed to FDRN and VoxelMorph and denoted respectively as FDRN$_{\downarrow\uparrow}$ and VoxelMorph$_{\downarrow\uparrow}$. The same scenario was performed for the upscaling of 4$\times$. Meanwhile, the trilinear upsampled images were also fed into ReFDRN and ReVoxelMorph and the results were compared with the ones of FDRN$_{\downarrow\uparrow}$ and VoxelMorph$_{\downarrow\uparrow}$. Besides, we performed image registration on the downscaled images and indicate them as FDRN$_{\downarrow}$ and VoxelMorph$_{\downarrow}$ to show the impact of image dimension change on the registration accuracy. The performance of the affine transformation was set as the baseline. The results of all the registrations for both upscaling factors are summarized in Table~\ref{tab:ReFDRN-ReVoxelMorph}. Comparing FDRN$_{\downarrow\uparrow}$ with ReFDRN$_{\downarrow\uparrow}$, VoxelMorph$_{\downarrow\uparrow}$ with ReVoxelMorph$_{\downarrow\uparrow}$, we can observe that image resolution does have noticeable impact on registration accuracy and the phenomenon becomes more evident when the input images are severely degraded such as in case of upscaling of 4$\times$. Comparing FDRN$_{\downarrow}$ with FDRN$_{\downarrow\uparrow}$ and VoxelMorph$_{\downarrow}$ with VoxelMorph$_{\downarrow\uparrow}$, it is obviously shown that not only resolution enhancement by SR, but also interpolation-based upscaling, which increases the image dimension, play an important role on the registration performance. 

\begin{figure*}
\centering
	\includegraphics[width=0.92\textwidth]{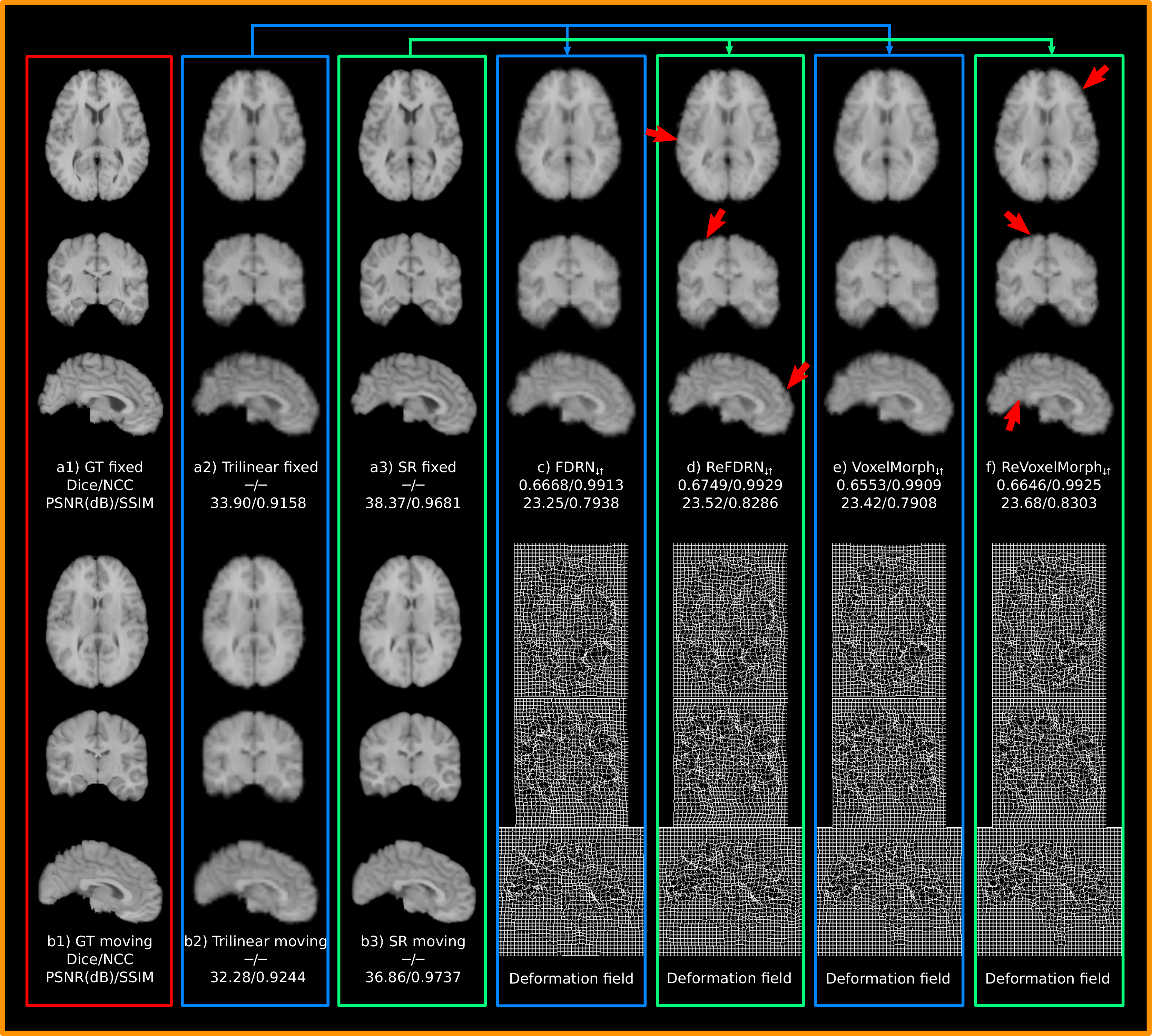}
	\caption{Visual evaluation of the impact of REM on the registration performance for upscaling factor of 4$\times$ on LPBA40. Red: GT; Blue: Trilinear interpolated images and the corresponding registration results; Green: SR images and the corresponding registration results.}
	\label{fig:VisualReFDRN}
\end{figure*}

In Fig.~\ref{fig:VisualReFDRN}, we demonstrate the visual comparison between the registration result based on trilinear interpolated images (marked in blue) and the one using SR outputs (marked in green) for the upscaling of 4$\times$. In particular, the GT images are shown in the very left column (marked in red) and they were downsampled as the LR inputs. The second and third columns depict the upscaled results of the downsampled inputs. One of the six upscaled testing images of LPBA40 was selected as the fixed image and the remaining five images were registered to it. The average of the five registered images by different methods are illustrated from column c) to column f) along with the deformation fields of one image pair and the corresponding Dice score, NCC, PSNR, and SSIM. Firstly, we can see that the SR results are much more pleasant containing sharper and finer structures than the trilinear interpolated ones. The superior performance of REM can also be verified by the quantitative evaluation where REM achieves an average performance gain of about 4.5dB in PSNR comparing to the trilinear interpolation. %Since the cascaded network ReFDRN and ReVoxelMorph produce not only the DVF but also the super-resolved images, the resolution enhanced images can be employed for further medical diagnosis. 
More importantly, the average registered images by ReFDRN$_{\downarrow\uparrow}$ and ReVoxelMorph$_{\downarrow\uparrow}$ are sharper than the ones by FDRN$_{\downarrow\uparrow}$ and VoxelMorph$_{\downarrow\uparrow}$ and resemble the fixed image more in visual perception. It indicates that the resolution enhanced networks by SR achieve a better registration accuracy and are more robust against different variants of the moving image. Besides, the quantitative assessment of the registration performance in Dice, NCC, PSNR, and SSIM coincides with the visual evaluation. In addition, the deformation fields of different networks from three views are exhibited and we do observe noticeable differences between the REM embedded cascaded networks and the interpolation-based naive ones.    

\begin{table}
	\centering
	\setlength{\tabcolsep}{8pt}
	\caption{Effectiveness of the introduced auxiliary loss at different upscaling factors. AL: auxiliary loss.}
	\label{tab:AL}
	\begin{threeparttable}
	\begin{tabular}{c  c  c  c  c  c}
		\toprule
		 Scales &Metrics& \multicolumn{2}{c}{ReFDRN} &\multicolumn{2}{c}{ReVoxelMorph} \\ \midrule
		 & &w/o AL &w/ AL &w/o AL&w/ AL \\ \cmidrule{3-6}
		\multirow{2}{*}{2$\times$}& Dice  & 0.6801 &0.6804 & 0.6730&0.6741 \\
								  & NCC  & 0.9974& 0.9977& 0.9965 &0.9970 \\ \midrule
		\multirow{2}{*}{4$\times$}& Dice & 0.6731 &0.6736 &0.6670 & 0.6676 \\
								  & NCC & 0.9959 &0.9962 &0.9923 &0.9932 \\ \bottomrule
	\end{tabular}
	\end{threeparttable}
\end{table}

\subsection{Evaluation of the Auxiliary Loss}\label{sec:evalauxiliary}
The effectiveness of the auxiliary loss formulated in \eqref{eq:Lossaux} is evaluated on the cascaded networks ReFDRN and ReVoxelMorph by Dice and NCC at different upscaling factors based on the LPBA40 dataset. For brevity, AL is short for the auxiliary loss. It is shown that on both networks and for both upscaling factors, the employment of the auxiliary loss does improve the registration performance in average Dice score and NCC.

\section{Conclusion}

In this work, we present a CNN-based resolution enhancement module (REM) and a cascaded resolution enhanced registration network. We demonstrate the impact of resolution enhancement on image registration quantitatively and qualitatively at different upscaling factors. In particular, different residual structures and network configurations of REM are evaluated on the LPBA40 brain MRI dataset. The proposed REM can be easily integrated into the registration network as a preprocessing module for spatial resolution enhancement. An auxiliary loss is introduced to the cascaded network to facilitate the robustness and enable a multi-hierarchical supervision. REM is evaluated on two registration networks FDRN and VoxelMorph for upscaling factors of 2$\times$ and 4$\times$. Experimental results demonstrate that REM improves the registration accuracy remarkably in Dice, NCC, PSNR, and SSIM especially when the input images are severely degraded with insufficient spatial resolution. In addition, along with the displacement vector field as the registration output, the cascaded network produces also the super-resolved images which can be further employed for medical diagnosis.

% if have a single appendix:
%\appendix[Proof of the Zonklar Equations]
% or
%\appendix  % for no appendix heading
% do not use \section anymore after \appendix, only \section*
% is possibly needed

% use appendices with more than one appendix
% then use \section to start each appendix
% you must declare a \section before using any
% \subsection or using \label (\appendices by itself
% starts a section numbered zero.)
%

% Can use something like this to put references on a page
% by themselves when using endfloat and the captionsoff option.
\ifCLASSOPTIONcaptionsoff
  \newpage
\fi

\bibliographystyle{IEEEtran}
\bibliography{egbib}

% Generated by IEEEtran.bst, version: 1.14 (2015/08/26)
\begin{thebibliography}{10}
\providecommand{\url}[1]{#1}
\csname url@samestyle\endcsname
\providecommand{\newblock}{\relax}
\providecommand{\bibinfo}[2]{#2}
\providecommand{\BIBentrySTDinterwordspacing}{\spaceskip=0pt\relax}
\providecommand{\BIBentryALTinterwordstretchfactor}{4}
\providecommand{\BIBentryALTinterwordspacing}{\spaceskip=\fontdimen2\font plus
\BIBentryALTinterwordstretchfactor\fontdimen3\font minus
  \fontdimen4\font\relax}
\providecommand{\BIBforeignlanguage}[2]{{%
\expandafter\ifx\csname l@#1\endcsname\relax
\typeout{** WARNING: IEEEtran.bst: No hyphenation pattern has been}%
\typeout{** loaded for the language `#1'. Using the pattern for}%
\typeout{** the default language instead.}%
\else
\language=\csname l@#1\endcsname
\fi
#2}}
\providecommand{\BIBdecl}{\relax}
\BIBdecl

\bibitem{survey1}
J.~A. Maintz and M.~A. Viergever, ``A survey of medical image registration,''
  \emph{Med. Image Anal.}, vol.~2, no.~1, pp. 1--36, 1998.

\bibitem{survey2}
J.~V. Hajnal, D.~L. Hill, and D.~J. Hawkes, ``Medical image registration,''
  \emph{Phys. Med. Biol.}, vol.~46, no.~3, 2001.

\bibitem{survey4}
A.~Sotiras, C.~Davatzikos, and N.~Paragios, ``Deformable medical image
  registration: A survey,'' \emph{IEEE Trans. Med. Imag.}, vol.~32, no.~7, pp.
  1153--1190, 2013.

\bibitem{survey5}
Y.~Fu, Y.~Lei, T.~Wang, W.~J. Curran, T.~Liu, and X.~Yang, ``Deep learning in
  medical image registration: a review,'' \emph{Phys. Med. Biol.}, 2020.

\bibitem{Beg2}
M.~F. Beg, M.~I. Miller, A.~Trouv\'{E}, and L.~Younes, ``Computing large
  deformation metric mappings via geodesic flows of diffeomorphisms,''
  \emph{Int. J. Comput. Vision}, vol.~61, no.~2, pp. 139--157, 2005.

\bibitem{Avants}
B.~B. Avants, C.~L. Epstein, M.~Grossman, and J.~C. Gee, ``Symmetric
  diffeomorphic image registration with cross-correlation: Evaluating automated
  labeling of elderly and neurodegenerative brain,'' \emph{Med. Image Anal.},
  vol.~12, no.~1, pp. 26--41, 2008.

\bibitem{Ashburner2}
J.~Ashburner and K.~J. Friston, ``Diffeomorphic registration using geodesic
  shooting and gauss-newton optimisation,'' \emph{NeuroImage}, vol.~55, no.~3,
  pp. 954--967, 2011.

\bibitem{dalca}
A.~V. Dalca, G.~Balakrishnan, J.~Guttag, and M.~R. Sabuncu, ``Unsupervised
  learning for fast probabilistic diffeomorphic registration,'' in \emph{Int.
  Conf. Med. Imag. Comp. Comput. Assist. Interv.}\hskip 1em plus 0.5em minus
  0.4em\relax Springer, 2018, pp. 729--738.

\bibitem{FDRN}
K.~Sun and S.~Simon, ``{FDRN}: A fast deformable registration network for
  medical images,'' \emph{Med. Phys.}, vol.~48, no.~10, pp. 6453--6463, 2021.

\bibitem{VoxelMorph}
G.~Balakrishnan, A.~Zhao, M.~R. Sabuncu, J.~Guttag, and A.~V. Dalca,
  ``Voxel{M}orph: a learning framework for deformable medical image
  registration,'' \emph{IEEE Trans. Image Process.}, vol.~38, no.~8, pp.
  1788--1800, 2019.

\bibitem{Bajcsy}
R.~Bajcsy and S.~Kova\v{c}i\v{c}, ``Multiresolution elastic matching,''
  \emph{Comput. Vis., Graph., Image Process.}, vol.~46, no.~1, pp. 1--21, 1989.

\bibitem{Gee}
J.~C. Gee and R.~Bajcsy, ``Elastic matching: Continuum mechanical and
  probabilistic analysis,'' \emph{Brain Warp.}, pp. 183--197, 1999.

\bibitem{Davatzikos}
C.~Davatzikos, ``Spatial transformation and registration of brain images using
  elastically deformable models,'' \emph{Comput. Vis. Image Understand},
  vol.~66, no.~2, pp. 207--222, 1997.

\bibitem{Rueckert}
D.~Rueckert, L.~I. Sonoda, C.~Hayes, D.~L.~G. Hill, M.~O. Leach, and D.~J.
  Hawkes, ``Nonrigid registration using free--form feformation: Application to
  breast mr images,'' \emph{IEEE Trans. Med. Imag.}, vol.~18, no.~8, pp.
  712--721, 1999.

\bibitem{Kybic}
J.~Kybic and M.~Unser, ``Fast parametric elastic image registration,''
  \emph{IEEE Trans. Image Process.}, vol.~12, no.~11, pp. 1427--1442, 2003.

\bibitem{Sdika}
------, ``A fast nonrigid image registration with constraints on the jacobian
  using large scale constrained optimization,'' \emph{IEEE Trans. Med. Imag.},
  vol.~27, no.~2, pp. 271--281, 2008.

\bibitem{Thirion}
J.~Thirion, ``Image matching as a diffusion process: an analogy with
  maxwell’s demons,'' \emph{Med. Image Anal.}, vol.~2, no.~3, pp. 243--260,
  1998.

\bibitem{Balakrishnan}
G.~Balakrishnan, A.~Zhao, M.~R. Sabuncu, J.~Guttag, and A.~V. Dalca, ``An
  unsupervised learning model for deformable medical image registration,'' in
  \emph{Proc. IEEE Conf. Comput. Vis. Pattern Recognit.}, 2018, pp. 9252--9260.

\bibitem{Beg1}
M.~F. Beg and A.~Khan, ``Symmetric data attachment terms for large deformation
  image registration,'' \emph{IEEE Trans. Med. Imag.}, vol.~26, no.~9, pp.
  1179--1189, 2007.

\bibitem{Huang}
T.~S. Huang and R.~Y. Tsan, ``Multiple frame image restoration and
  registration,'' \emph{Advances in Computer Vision and Image Process., JAI
  Press, Inc. Greenwich, CT}, pp. 317--339, 1984.

\bibitem{Kim2}
S.~P. Kim and W.~Su, ``Recursive high-resolution reconstruction of blurred
  multiframe images,'' \emph{IEEE Trans. Image Process.}, vol.~2, no.~4, pp.
  534--539, 1993.

\bibitem{BTV}
S.~Farsiu, M.~D. Robinson, M.~Elad, and P.~Milanfar, ``Fast and robust
  multiframe super-resolution,'' \emph{IEEE Trans. Image Process.}, vol.~13,
  no.~10, pp. 1327--1344, 2004.

\bibitem{Elad}
M.~Elad and Y.~Hel-Or, ``A fast super-resolution reconstruction algorithm for
  pure translational motion and common space-invariant blur,'' \emph{IEEE
  Trans. Image Process}, vol.~10, no.~8, pp. 1187--1193, 2001.

\bibitem{Hardie}
R.~Hardie, K.~Barnard, and E.~Armstrong, ``Joint map registration and
  high-resolution image estimation using a sequence of undersampled images,''
  \emph{IEEE Trans. Image Process.}, vol.~6, no.~12, pp. 1621--1633, 1997.

\bibitem{IRWSR}
T.~K{\"o}hler, X.~Huang, F.~Schebesch, A.~Aichert, A.~Maier, and J.~Hornegger,
  ``Robust multiframe super-resolution employing iteratively re-weighted
  minimization,'' \emph{IEEE Trans. Comput. Imag.}, vol.~2, no.~1, pp. 42--58,
  2016.

\bibitem{MPG}
K.~Sun, T.~Tran, R.~Krawtschenko, and S.~Simon, ``Multi-frame super-resolution
  reconstruction based on mixed {P}oisson--{G}aussian noise,'' \emph{Signal
  Process. Image Commun.}, vol.~82, p. 115736, 2020.

\bibitem{MPGBSWTV}
K.~Sun and S.~Simon, ``Bilateral spectrum weighted total variation for
  noisy-image super-resolution and image denoising,'' \emph{IEEE Trans. Signal
  Process.}, vol.~69, pp. 6329--6341, 2021.

\bibitem{FLMISR}
K.~Sun, T.~Tran, J.~Guhathakurta, and S.~Simon, ``Fl-misr: fast large-scale
  multi-image super-resolution for computed tomography based on multi-gpu
  acceleration,'' \emph{J. Real-Time Image Process.}, pp. 1--14, 2021.

\bibitem{SRCNN}
C.~Dong, C.~C. Loy, K.~He, and X.~Tang, ``Learning a deep convolutional network
  for image super-resolution,'' in \emph{Proc. Eur. Conf. Comput. Vis.}, 2014,
  pp. 184--199.

\bibitem{SRGAN}
C.~Lediga \emph{et~al.}, ``Photo-realistic single image super-resolution using
  a generative adversarial network,'' in \emph{Proc. IEEE Conf. Comput. Vis.
  Pattern Recognit.}, 2017, pp. 105--114, [doi: 10.1109/CVPR.2017.19].

\bibitem{VDSR}
J.~Kim, J.~K. Lee, and K.~M. Lee, ``Accurate image super-resolution using very
  deep convolutional networks,'' in \emph{Proc. IEEE Conf. Comput. Vis. Pattern
  Recognit.}, 2016, pp. 1646--1654.

\bibitem{EDSR}
B.~Lima, S.~Son, H.~Kim, S.~Nah, and K.~M. Lee, ``Enhanced deep residual
  networks for single image super-resolution,'' in \emph{Proc. IEEE Conf.
  Comput. Vis. Pattern Recognit. Workshops}, 2017, pp. 136--144.

\bibitem{DPSR}
K.~Zhang, W.~Zuo, and L.~Zhang, ``Deep plug-and-play super-resolution for
  arbitrary blur kernels,'' in \emph{Proc. IEEE Conf. Comput. Vis. Pattern
  Recognit.}, 2019, pp. 1671--1681.

\bibitem{ESRGAN}
X.~Wang \emph{et~al.}, ``{ESRGAN}: Enhanced super-resolution generative
  adversarial networks,'' in \emph{Proc. Eur. Conf. Comput. Vis.}, 2018, pp.
  1--16.

\bibitem{DaiD}
D.~Dai, Y.~Wang, Y.~Chen, and L.~Van~Gool, ``Is image super-resolution helpful
  for other vision tasks?'' in \emph{IEEE Winter Conf. Applications Comput.
  Vis.}, 2016, pp. 1--9.

\bibitem{dualsr}
L.~Wang, D.~Li, Y.~Zhu, L.~Tian, and Y.~Shan, ``Dual super-resolution learning
  for semantic segmentation,'' in \emph{Proc. IEEE Conf. Comput. Vis. Pattern
  Recognit.}, 2020, pp. 3774--3783.

\bibitem{dualface}
P.~H. Hennings-Yeomans, S.~Baker, and B.~V. Kumar, ``Simultaneous
  super-resolution and feature extraction for recognition of low-resolution
  faces,'' in \emph{Proc. IEEE Conf. Comput. Vis. Pattern Recognit.}\hskip 1em
  plus 0.5em minus 0.4em\relax IEEE, 2008, pp. 1--8.

\bibitem{srreid}
X.~Jing \emph{et~al.}, ``Super-resolution person re-identification with
  semi-coupled low-rank discriminant dictionary learning,'' in \emph{Proc. IEEE
  Conf. Comput. Vis. Pattern Recognit.}, 2015, pp. 695--704.

\bibitem{SRDenseNet}
T.~Tong, G.~Li, X.~Liu, and Q.~Gao, ``Image super-resolution using dense skip
  connections,'' in \emph{Proc. IEEE Int. Conf. Comput. Vis.}, 2017, pp.
  4799--4807.

\bibitem{rcan}
Y.~Zhang \emph{et~al.}, ``Image super-resolution using very deep residual
  channel attention networks,'' in \emph{Proc. Eur. Conf. Comput. Vis.}, 2018,
  pp. 286--301.

\bibitem{LPBA40}
D.~W. Shattuck \emph{et~al.}, ``Construction of a 3d probabilistic atlas of
  human cortical structures,'' \emph{Neuroimage}, vol.~39, no.~3, pp.
  1064--1080, 2008.

\bibitem{unbiased}
V.~Fonov \emph{et~al.}, ``Unbiased average age-appropriate atlases for
  pediatric studies,'' \emph{Neuroimage}, vol.~54, no.~1, pp. 313--327, 2011.

\end{thebibliography}

\end{document}